\documentclass[runningheads]{llncs}

\usepackage{algorithmic}
\usepackage{graphicx}
\usepackage{textcomp}
\usepackage{xcolor}

\usepackage[utf8]{inputenc}
\usepackage{lmodern}
\usepackage[english]{babel}
\usepackage{graphicx}
\usepackage{xspace}
\usepackage{tabulary}
\usepackage{booktabs,multirow,tabularx}
\usepackage{hyperref}
\usepackage{subcaption}
\usepackage{harveyballs}
\usepackage{booktabs}

\usepackage[noabbrev,capitalise]{cleveref}

\usepackage{xcolor}

\newcommand{\explicitAAS}{
    \def\harveyBallsColor{cyan}
    \def\harveyBallsLineColor{cyan}
    \harveyBallFull{}
}
\newcommand{\partialAAS}{
    \def\harveyBallsColor{cyan}
    \def\harveyBallsLineColor{cyan}
    \harveyBallHalf{}
}
\newcommand{\implicitAAS}{
    \def\harveyBallsColor{gray}
    \def\harveyBallsLineColor{gray}
    \reflectbox{\harveyBallHalf}
}
\newcommand{\nosupportAAS}{
    \def\harveyBallsColor{gray}
    \def\harveyBallsLineColor{gray}
    \harveyBallNone{}
}
\begin{document}
%
\title{A Digital Twin Description Framework and its Mapping to Asset Administration Shell}

\titlerunning{A Digital Twin Description Framework and Mapping to AAS}
%
\author{Bentley James Oakes\inst{1,2}\thanks{B. Oakes carried out the majority of this work at the University of Antwerp.}\orcidID{0000-0001-7558-1434} \and
	Ali Parsai\inst{3}\thanks{A. Parsai is now employed at Agfa Offset \& Inkjet Solutions, Mortsel, Belgium.}\orcidID{0000-0001-8525-8198} \and
	Bart Meyers\inst{3}\orcidID{0000-0001-9566-8297} \and
	Istvan David\inst{1,4}\orcidID{0000-0002-4870-8433} \and
	Simon Van Mierlo\inst{3,5}\thanks{S. Van Mierlo is now employed at EP\&C Patent Attorneys, Turnhout, Belgium.}\orcidID{0000-0002-4043-6883} \and
	Serge Demeyer\inst{5}\orcidID{0000-0002-4463-2945} \and
	Joachim Denil\inst{3,5}\orcidID{0000-0002-4926-6737} \and 
	Paul De Meulenaere\inst{3,5}\orcidID{0000-0002-3706-6164} \and
	Hans Vangheluwe\inst{3,5}\orcidID{0000-0003-2079-6643}}
\authorrunning{B. Oakes et al.}
%
\institute{Polytechnique Montr{\'e}al, Montr{\'e}al, Canada\\
	\email{bentley.oakes@polymtl.ca} \and
	Universit{\'e} de Montr{\'e}al, Montr{\'e}al, Canada\\ \email{\{bentley.oakes, istvan.david\}@umontreal.ca}   \and
	Flanders Make vzw, Lommel, Belgium\\ \email{\{ali.parsai,bart.meyers\}@flandersmake.be} \and
	McMaster University, Hamilton, Canada\\
	\email{davidi3@mcmaster.ca} \and
	University of Antwerp, Antwerp, Belgium\\ \email{\{serge.demeyer, joachim.denil, paul.demeulenaere, hans.vangheluwe\}@uantwerpen.be}}
\maketitle              

\noindent\textcolor{red}{The final authenticated version is available online at \url{https://doi.org/10.1007/978-3-031-38821-7_1}.}
%
%
%
\abstract{The pace of reporting on Digital Twin (DT) projects continues to accelerate both in industry and academia. However, these experience reports often leave out essential characteristics of the DT, such as the scope of the system-under-study, the insights and actions enabled, and the time-scale of processing. A lack of these details could therefore hamper both understanding of these DTs and development of DT tools and techniques.
Our previous work developed a DT description framework with fourteen characteristics as a checklist for experience report authors to better describe the capabilities of their DT projects. This report provides an extended example of reporting to highlight the utility of this description framework, focusing on the DT of an industrial drilling machine. Furthermore, we provide a mapping from our description framework to the Asset Administration Shell (AAS) which is an emerging standard for Industry 4.0 system integration. This mapping aids practitioners in understanding how our description framework relates to AAS, potentially aiding in description or implementation activities.}

\section{Introduction}
\label{sec:intro}

The \textit{digital twinning} concept is now prevalent in multiple domains and industries~\cite{Rasheed2020,Jones2020}. This is due to \textit{digital twins} (DTs) allowing system designers, manufacturers, business stakeholders, and other users to explore possibilities in digital versions of their \textit{system-under-study} (SUS). 

For a useful definition of DTs, we point to Madni \textit{et al.}, who state ``a DT is a virtual instance of a physical system (twin) that is continually updated with the latter’s performance, maintenance, and health status data throughout the physical system’s life-cycle''~\cite{Madni2019}. This is an expanded definition from the original of Grieves \textit{et al.}, who focused on \textit{product life-cycle management} where the DT represented either the pre-manufactured product or the product in usage~\cite{Grieves2017}.

These DTs can be at multiple scales, such as monitoring air quality with a few sensors~\cite{Govindasamy2021}, representing individual machines in a factory~\cite{Min2019}, or monitoring the energy management of an entire district in Helsinki~\cite{Ruohomaeki2018}. These DTs can be implemented to reason about the behaviour of a SUS in the past, present, or future in various conditions, allowing for unprecedented exploration of a system's dynamics. For example, usages of DTs can include automatic scheduling of maintenance~\cite{Werner2019}, anomaly detection and prediction, visualisation, and system optimisation~\cite{Rasheed2020}.


\paragraph{Describing Digital Twins}

Our previous work has shown that both academic and industrial experience reports omit crucial information, such as the time-scales or automatic nature of activities~\cite{Oakes2021} This leads to confusion about the capabilities and classification of the DTs.

For example, Kritzinger \textit{et al.} define three categories of DT: \textit{digital model} (DM), \textit{digital shadow} (DS), and \textit{digital twin} (DT). The criteria is whether the communication between the SUS and the DT is manual or automatic. In a \textit{digital model}, data is not automatically sent from a SUS to a DT, and any \textit{actions} from the DT to the SUS are manually performed. A \textit{digital shadow} has an automatic data connection from the SUS to the DT, and a \textit{digital twin} (as defined by Kritzinger \textit{et al.}) has automatic transfer of data and automated commands from the DT to the SUS. 

Our earlier paper showed that even this simple classification cannot be determined in some experience reports, leading to uncertainty about the capabilities of the DT solution~\cite{Oakes2021}. The example in our earlier paper is a DT for a ``human-robot collaborative work environment''~\cite{Malik2018}, where robot control instructions are generated to prevent collisions between the robot and their human co-worker. However, it is unclear whether any code or instructions are automatically uploaded to the robot when analysis is performed. Thus, it is unclear whether the report describes a \textit{digital shadow} or a \textit{digital twin}.

To address this issue of imprecisions in experience reports, our earlier work suggests fourteen characteristics to describe in experience reports about DTs~\cite{Oakes2021}. This structured approach allows for greater insight and clarity about the capabilities of DTs and their development. In particular, we wish authors to clarify their expectations about the term ``digital twin'', whether it is real-time control~\cite{Zhuang2018}, an enhanced tracking simulator~\cite{Werner2019}, or a high-fidelity model~\cite{Miller2018}. Five experience reports from the literature are presented with these fourteen characteristics, with a further fifteen reports in an online table~\cite{repo}. Of particular interest is that we found six reports where the classification suggested by our characteristics differs from those of \cite{Kritzinger2018} and \cite{Fuller2019}. 

\paragraph{Paper Contributions and Structure}

In this work, we further expand the presentation and applicability of our earlier paper. In particular, we present a DT of an industrial drilling machine with ``smart clamp'' suction cups~\cite{Bey-Temsamani2019} in \cref{sec:use_case}. The capabilities of this DT is expressed using the characteristics of our description framework~\cite{Oakes2021} in \cref{sec:framework}. This assists with the understanding of each characteristic and provides an example of their use in understanding a DT's capabilities. We also add an relevant detail to our list concerning whether a usage of a DT focuses on mainly \textit{historical} information from the past of a SUS, on \textit{streaming/live} information from the present time, or both~\cite{Oakes2021b}.

To demonstrate the utility of our framework, we map it onto the Asset Administration Shell (AAS) in \cref{sec:aas}. The AAS provides standardized techniques to describe digital assets in a hierarchical fashion, but mostly focuses on the lower-level implementation details such as data and functions. In contrast, our framework enables describing high-level capabilities in a less formal, narrative-based fashion. Mapping our framework onto AAS serves two reasons: a) to allow authors who have a DT implemented in AAS to better express the characteristics by our description framework; and b) to offer a high-level starting point for AAS concepts.

\Cref{sec:conclusion} then concludes and provides directions for future work.

\section{Running Example}
\label{sec:use_case}

This section describes the running example for this paper: an industrial drilling machine, augmented with a Digital Twin (DT) to improve drilling performance, monitor tool wear, and provide real-time feedback on the machine's operation. This complex cyber-physical system has been developed as a demonstrator and research platform by Flanders Make~\footnote{\url{https://www.flandersmake.be/en}}, the strategic research center for the Flemish manufacturing industry.

\paragraph{First Project Phase: Smart Clamp Drilling}

As described by Bey-Temsamani \textit{et al.}~\cite{Bey-Temsamani2019} the impetus behind the \textit{smart clamp} project was to investigate innovations for improving drilling performance in composite materials which are useful for applications such as aeronautics. When the drill is forced against the plate and when exiting the material, there are significant forces applied to the plate due to the strength of the composite materials and the plate will move (\textit{deflect}). This can ruin the surface coating (lamination) and circularity of the hole which directly affects the ability for plates to be fastened together.

\begin{figure}
	\centering
	\includegraphics[width=0.55\linewidth]{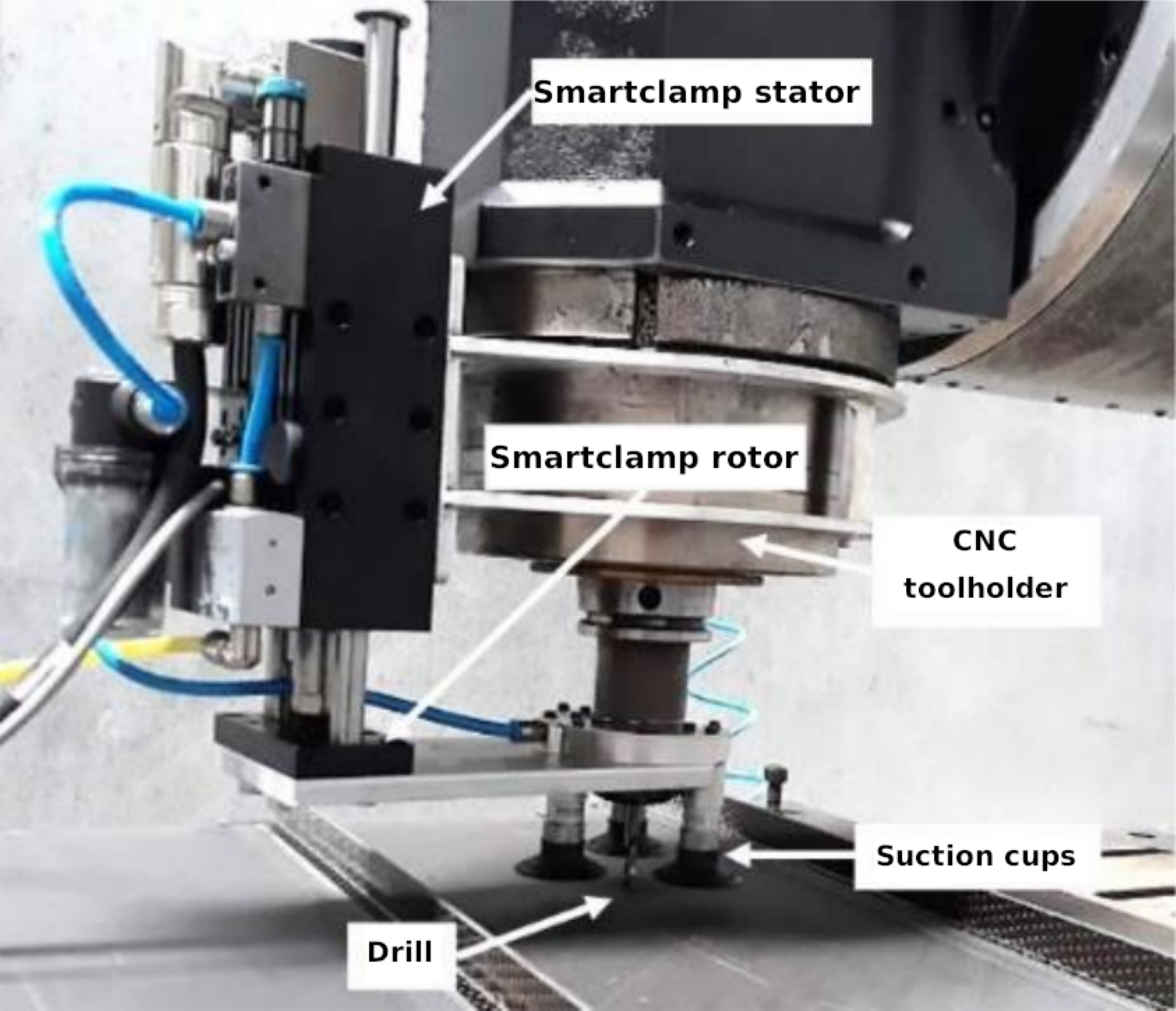}
	\caption{A labelled photo of the drilling machine with ``smart clamp'' suction cups to secure the plate during drilling~\cite{Bey-Temsamani2019}.}
	\label{fig:smart_clamp_pic}
\end{figure}

Clamps can be used to secure the plate during drilling, preferably custom clamps designed for each piece to be secured. However, the clamps are time-consuming to construct and attach, and they must be very close ($<$~80 millimetres) to the drill to properly secure the plate~\cite{Bey-Temsamani2019}. Therefore, the first improvement developed by Bey-Temsaman \textit{et al.} is a patented \textit{smart clamp} that uses suction cups which compensate in real-time for the drill and plate motion during drilling~\cite{SmartClampPatent}. To reduce plate deflection during drilling, the suction cups adjust the position of the plate in a real-time control loop as pictured in Figure~\ref{fig:smart_clamp_pic}.

The second innovation in the smart clamp demonstrator is a method for detecting tool wear using an optical one-dimensional sensor. This sensor measures the wear of the drill bit directly by measuring the dimensions of the drill to test against the original bit, allowing for replacement only when necessary to maintain performance. Before the drill bit has degraded enough to impact the quality of the hole, the user can decide that a bit change is required. This ensures hole quality while avoiding unnecessary replacement of the drill bit.

\paragraph{Second Project Phase: Reporting on Quality Metrics}

Following the initial successes on the smart clamp, other projects at Flanders Make have used the smart clamp as a test-bed platform for further innovations. For example, the smart clamp platform has been extended with an Internet of Things (IoT) architecture to enable the reporting and storage of drill and hole quality metrics.

Specifically, the storage and retrieval of historical data was implemented such that the control algorithm could be improved and further correlations detected. Metrics of the drilling process are processed and sent to a visualisation dashboard for the user. This includes the thickness of the plate, the hole location, a picture of the drilled hole, and the results from a vision algorithm for detecting plate deflection during drilling.



\section{Digital Twin Description Framework}
\label{sec:framework}

This section expands on the description framework from our earlier work~\cite{Oakes2021} which aims to precisely describe digital twins (DTs), their system-under-study (SUS), and the nature of their relation. We focus on presenting DTs as a \textit{constellation} of supporting components to support a \textit{usage} for that DT, with the constellation evolving over time to support further usages. These characteristics encourage authors of experience reports to report the capabilities of their DTs in appropriate details, so that they can be correctly understood and classified.



\subsection{Summary of Characteristics}
\label{sec:key_char}

\begin{figure}[tbph]
	\begin{center}
		\includegraphics[width=\textwidth]{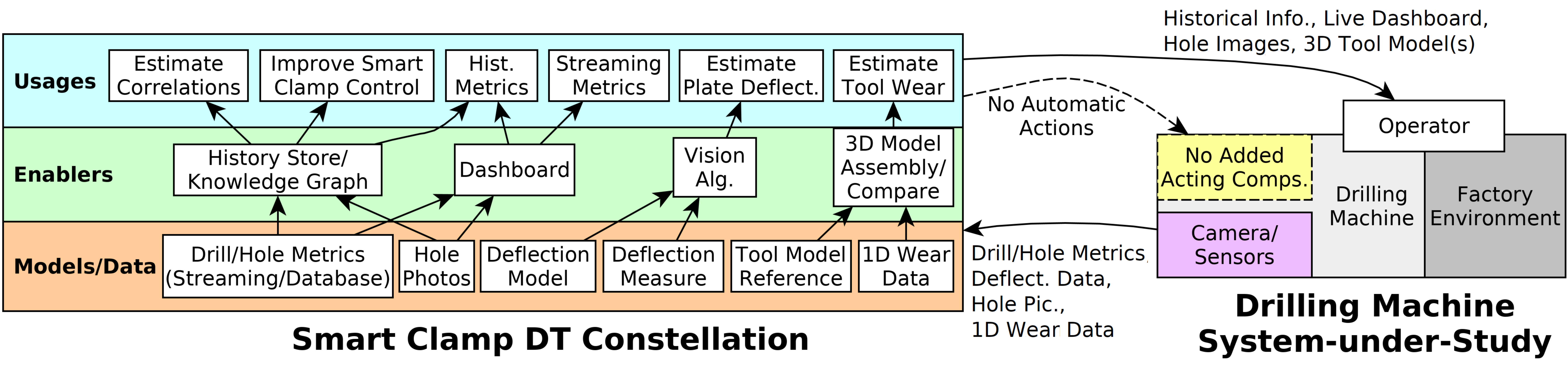}
	\end{center}
	\vspace{-20px} 
	\caption{The smart clamp DT represented in our description framework.}
	\label{fig:scre}
\end{figure}

This section summarises the ``smart clamp'' drilling machine DT discussed in \cref{sec:use_case} with the fourteen characteristics we have selected for our description framework, as labelled from \textit{C1} to \textit{C14}. The descriptions here are intentionally brief and further information can be found throughout \cref{sec:use_case} and \cref{sec:framework}. However, we hope that these brief lines, along with \Cref{fig:scre} illustrating the relationships, can allow readers to understand the purpose and utility of the smart clamp DT.



Note that the smart clamp system is actually a \textit{Digital Shadow} by the classification of Kritzinger \textit{et al.}~\cite{Kritzinger2018}. This is due to the lack of automatic control from the DT on the drilling machine as reported by the \textit{C6: Insights and Actions characteristic}. This lack of automatic control is a deliberate one as it common in industrial requirements to always have the machine operator in the loop. Thus, information is provided to the operator but no automatic actions are directly performed by the DT.

\paragraph{C1: System-under-study} - \cref{sec:fw_sus} - \textit{The scope of the SUS.}

\textit{System:} Drilling machine with smart clamps and plates.

\textit{Environment:} Surroundings including temp. and humidity.

\textit{Agent:} Drilling machine operator.


\paragraph{C2: Acting Components} - \cref{sec:fw_act_sense} - \textit{Additions and modifications to the SUS enabling DT actions and insights.}

Hardware to store and display dashboard metrics.

\paragraph{C3: Sensing Components} - \cref{sec:fw_act_sense} - \textit{Additions and modifications to the SUS enabling DT data collection.}

Camera to capture hole photos, infrastructure to send data to history store and dashboard.

\paragraph{C4: Multiplicities} - \cref{sec:fw_multiplicity} - \textit{How many DTs and SUS entities are involved in the solution, and their relationship.}

One drilling machine connected to DT instances for each usage.

\paragraph{C5: Data Transmitted} - \cref{sec:fw_data_insights_actions} - \textit{Info. from SUS to DT.}

\textit{Manual:} None. \textit{Automatic:} Metrics on motor load, deflection reading, hole metrics and picture, tool dimensions.

\paragraph{C6: Insights/Actions} - \cref{sec:fw_data_insights_actions} - \textit{Info./control from DT to SUS.}

\textit{Insights:} Drill performance correlations, tool wear, machine/hole metrics.

\textit{Manual Actions:} Adjustment of drilling parameters, changing tool bit.

\textit{Automatic Actions:} None.

\paragraph{C7: Usages} - \cref{sec:fw_usages} - \textit{The activities the DT is used for.}

Estimate correlations, improve the smart clamp control, historical and streaming metrics for drill and holes, estimate plate deflection, and estimate tool wear.

\paragraph{C8: Enablers} - \cref{sec:fw_enablers} - \textit{The DT components which use models and data to support usages.}

A historical store (or \textit{knowledge graph}), a dashboard for the operator, vision algorithm, and three-dimensional model comparison.

\paragraph{C9: Models and Data} - \cref{sec:fw_models_and_data} - \textit{Input/output for enablers.}

Streaming and historical metrics for the drill and holes, photos of the holes, a model for calculating deflection and the incoming measurement, a reference model for the current tool and the incoming tool dimensions.

\paragraph{C10: Constellation} - \cref{sec:fw_constellations} - \textit{Relationships between usages, enablers, models/data.}

\Cref{fig:scre} shows the constellation of the smart clamp DT, with relationships between components shown by arrows.

\paragraph{C11: Time-Scale} - \cref{sec:fw_time_scales} - \textit{The time-scale of the data, insights, actions, and simulations used.}

\textit{Slower-than-real-time:} Find correlations, improve smart clamp control.

\textit{Real-time:} Dashboard updates to operator, storage in historical store. 

\textit{Faster-than-real-time:} None.

\paragraph{C12: Fidelity Considerations} - \cref{sec:fw_high_fidelity} - \textit{Explanations of fidelity of DT to SUS with respect to each DT usage.}

Moderate demands due to noisy data from manufacturing environment and resolution of sensors. Tool wear reasoning is more tolerant due to gradual decline and compensation for sensor contamination.

\paragraph{C13: Life-cycle Stages} - \cref{sec:fw_life_stages} - \textit{Life-cycle stages the DT is utilized for, usages for each, and (if varying) the scope of the SUS.}

\textit{Design:} Estimate correlations, improve smart clamp control, historical metrics.

\textit{Operation:} Improve smart clamp control, stream metrics, estimate plate deflection, and estimate tool wear.

\paragraph{C14: Evolution} - \cref{sec:fw_evolution} - \textit{How the DT evolves during development.}

Correlations found, smart clamp built and programmed, deflection and tool wear sensors developed, then dashboard built.



\subsection{Relating Digital Twin(s) and System-under-Study}

\begin{figure}[tbh]
	\centering
	\includegraphics[width=0.70\linewidth]{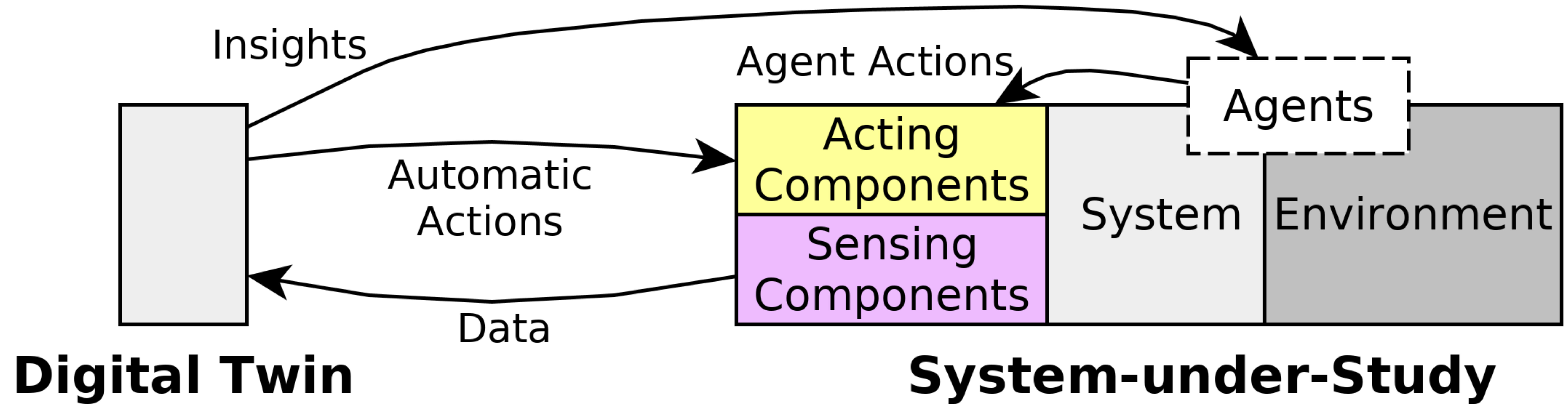}
	\caption{Digital Twin and the System-Under-Study (replicated from~\cite{Oakes2021}).}
	\label{fig:dt_sus}
\end{figure}

The relationship between the DT and the SUS is at the core of the DT concept. This is due to the ``twinning'' of the information of the SUS within the DT, as well as the communication from the DT back to the SUS. In \cref{fig:dt_sus}, this relationship is pictured with the DT as a black-box system which is examined further in \cref{sec:fw_layers}.

\subsubsection{C1: System-Under-Study}
\label{sec:fw_sus}

Within the theory of modelling and simulation, the SUS takes prime importance for a practitioner to reason about~\cite{Zeigler2000}. This is due to the critical requirement for a practitioner to understand the bounds, influences, and properties of the SUS. This is a highly non-trivial task to clearly define the boundaries of a system, and an author of an experience report must take care to precisely identify the relevant components of their study.

Our
framework takes the SUS to include the primary interacting entities (the \textit{system}), as well as the context (or \textit{environment}) surrounding and interacting with those entities. In the drilling machine example, the actual system is the machine with its software, signals, mechanical components, and the composite plate, while the environment includes the surrounding air pressure, temperature, humidity, ground vibrations, etc.

It is also important to denote the \textit{agents} interacting with the SUS. In \cref{fig:dt_sus}, these agents are represented by a dashed extension box, as they may be considered part of the SUS, or an external force acting on the SUS. An example of such an agent would be a human operator who directly manipulates the drilling machine in our running example, or healthcare agents as part of a healthcare system~\cite{Liu2019}. However, our DT description framework also allows for artificial intelligence (AI) agents to be considered a part of the SUS. This is relevant for other DTs found in the literature such as the AI agents described in~\cite{Wuest2015}.

Finally, \cref{fig:dt_sus} conceptually separates the DT and the SUS into two distinct components for understandability. However, in the systems involved there may not be such a clean physical separation. For example, sensing, processing, or acting components required for the operation of the DT may be physically present on or within the SUS. These components can directly influence the system through logical effects (competing for processing/memory/communication resources), or physical effects (temperature, vibration). Therefore, our description framework is only a starting point for an author of an experience report to more precisely define what they believe to be the SUS and what is the DT.

\paragraph{Smart Clamp DT:}

For the smart clamp example, a crucial question is whether the smart clamp controller is considered part of the underlying drilling system, or part of the DT. For the purposes of this report, the smart clamp functionality focusing on sensing the drilling procedure and controlling of the clamping system is considered part of the drilling machine (the SUS). This decision was made based on the fact that the smart clamp controller does not store any history of the drilling machine, but only reacts based on incoming information. Thus, the part of the smart clamp system which is included in our DT example primarily focuses on the communication to the user, such as dashboards, history storage, and reporting on tool and hole metrics.

The \textit{system} is considered to be the drilling machine with smart clamps. This includes the material plates, drill and bit, drilling forces, the clamping mechanism, clamp controller, drill (CNC) controller and control algorithm. The \textit{environment} for the drilling machine is the surroundings of the drilling machine, including any influence from ambient temperature and vibration. Finally, the \textit{agent} in the SUS is the machine operator, who externally manipulates the SUS but is not otherwise modelled.


\subsubsection{C2 and C3: Acting and Sensing Components}
\label{sec:fw_act_sense}

As practitioners develop DTs for their SUS, it may become necessary to add or modify components on the SUS to support uni- or bi- directional communication between the DT and the SUS~\cite{Chhetri2019,Lindstroem2017}. A common example is the addition of Internet of Things (IoT) sensors or communication hardware to transfer data. We propose that understanding this SUS modification process is crucial to reason about the cost and effort for building DTs of a SUS. Thus, we ask that authors report the (interesting) modifications of either \textit{acting} or \textit{sensing} SUS components.


\textit{Acting components} are those that permit \textit{actions} on the SUS by the DT. That is, actions are received from the DT or by agents, and some state change is effected on the SUS. This may be a physical actuator to operate a switch on the SUS, or a component such as a Programmable Logic Controller able to modify software parameters.

\textit{Sensing components} are those which collect and transmit information from the SUS to the DT. Typically, these are IoT components, such as a motor load sensor in the smart clamp SUS. Experience report authors may also wish to include larger components such as the addition of a Product Life-cycle Management system and related hardware to store product data in this category~\cite{Tao2018}. However, these systems and components are more likely part of the DT, outside of the scope of the SUS. Thus, they could be mentioned in a supporting technology discussion, which we do not consider in our list of fourteen characteristics.

The
\textit{acting} and \textit{sensing} components described here are left intentionally underspecified for the authors of an experience report to adapt them to their DT example. We encourage authors to answer the following questions with these characteristics, \textit{what was added to the SUS to enable communication with the DT}, and \textit{what was added to the SUS to enable control by the DT}. Answering these questions will allow practitioners and researchers to better understand and categorize the development of DTs.

\paragraph{Smart Clamp DT:}

For the smart clamp system, nothing was added to the SUS to enable control by the DT, due to no desire for automatic control. For enabling insights, hardware has been added nearby the SUS to display a dashboard to the operator.
A multitude of sensing components have been added, however. These include a camera to capture photos of the drilled hole quality, and the infrastructure to communicate this image and other hole metrics back to digital storage and the dashboard.

\subsubsection{C4: Multiplicity}
\label{sec:fw_multiplicity}

An assumption a reader may have while reading a DT experience report is that there is a one-to-one relationship of a DT to a SUS. However, this may not be true. We consider in our description framework that one SUS may be connected to multiple DTs (see \cref{sec:fw_constellations}). Each one of these DTs (called a \textit{DT instance}) then handles one particular \textit{usage}, by receiving and/or sending \textit{data}, \textit{insights}, and \textit{actions} (see \cref{sec:fw_data_insights_actions}) to and from the SUS.

The SUS itself may also be composed of multiple entities. For example, consider a manufacturing factory with an array of machines. A DT instance could be constructed for each individual machine to receive data from that machine and provide insights and actions for that particular machine. This group of DTs each connected to one particular machine is then termed as a \textit{DT Aggregate} by \cite{Grieves2017}. An alternative is to construct a DT for the factory itself which receives data for each machine (or statistical measures for the group) and insights or actions are sent for the conceptual collective. The decision about which architecture to implement is based on the system design implemented by the practitioner, as well as any technical restrictions on distributed computing.

It may not be obvious how many DTs are communicating with one particular SUS, and what the organization of entities is within the SUS. Our description framework therefore suggests that an experience report explicitly describe the \textit{multiplicity} of the SUS and the DTs and discuss the pattern of communication within the SUS and to each DT.

\paragraph{Smart Clamp DT:}

In the running example, there is one solitary drilling machine as the SUS, with one motor, drill, drill bit, and a single plate during each drilling operation. Of course, there are multiple holes drilled during the drilling process. Connected to this one SUS are DT instances (see \cref{sec:fw_constellations}) to provide insights on the hole quality, and tool wear. That is, a conceptual DT instance is used for insights on hole quality, and another one for tool wear.

\subsection{Information Connection}
\label{sec:fw_connection}

The backbone of the \textit{digital twinning} activity is the information connection between the SUS and the DT~\cite{Grieves2017}, where the circular and continuous flow of information allows the DT to mirror the SUS~\cite{Worden2020}. As
shown in \cref{fig:dt_sus}, changes in the SUS are propagated to the DT to be reflected, and actions and insights flow from the DT back towards the SUS.


We also recall here for the reader the useful classification of Kritzinger \textit{et al.}~\cite{Kritzinger2018}, which separates \textit{digital representations} into \textit{digital models}, \textit{digital shadows}, and \textit{digital twins}. This classification depends on the automatic nature of this connection. That is, a lack of automatic information flow from the SUS towards the digital representation means that it is a \textit{digital model}. If there is no automatic information flow (automatic actions) from the digital representation to the SUS, then the digital representation is a \textit{digital shadow}, also called a \textit{tracking simulator}. It is only with an automatic connection in both directions that the digital representation is denoted as a \textit{digital twin}. Again, the smart clamp DT does not have automatic control on the SUS, thus it is a \textit{digital shadow}.

\subsubsection{C5 and C6: Data, Insights, and Actions}
\label{sec:fw_data_insights_actions}

One of the most crucial questions to be asked about a DT is \textit{what does it do?}. In our description framework for reporting on DTs, we break this down into three broad categories: \textit{data}, \textit{insights}, and \textit{actions}. These categories specify what information is passed back and forth between the SUS and the DT, and are extremely useful for understanding their relationship.

\textit{Data} is specified as any information which flows from the SUS to the DT for processing or storage. As mentioned above, this flow may be manual (for a \textit{digital model}) or automatic (for a \textit{digital shadow/twin}).

\textit{Insights} are actionable pieces of information travelling from the DT towards the SUS. However, insights do not provoke a change in the SUS directly. Instead, they are transmitted to the agents surrounding the SUS, who may then decide to modify the SUS. Examples in this category include dashboards on the SUS's performance, alerts about unexpected behaviour, or reports indicating a potential for improvement. For instance, Zhuang \textit{et al.} implemented a DT which simulated a factory's geometry and the behaviour of the workers to provide insights on improved layout possibilities~\cite{Zhuang2018}.

\textit{Actions} are divided into two categories: \textit{automatic actions} and \textit{agent actions}. \textit{Automatic actions} are those commands transmitted by the DT to the SUS, which directly provoke a change in the SUS. An example of an automatic action is a control signal from the DT to adjust parameters in the SUS~\cite{Min2019,Ludvigsen2016}. In contrast, \textit{agent actions} are the actions which agents can modify the system, either as a physical or digital action. According to the classification of Kritzinger \textit{et al.}~\cite{Kritzinger2018}, there must be \textit{automatic actions} for a digital representation to be classified as a \textit{digital twin}.

\paragraph{Smart Clamp DT:}

For the smart clamp, a number of metrics need to be communicated as \textit{data} to the DT for displaying to the operator and storage in a historical database. These include a measure of motor load on the drill, a sensor reading to calculate deflection of the plate during drilling, various metrics of the drilled hole, and a picture of the hole. The tool wear also creates a three-dimensional model of the drill bit for inspection by the operator who decides on replacing the bit~\cite{Bey-Temsamani2019}.

A number of insights are provided to the machine operator for their consideration through both a dashboard for \textit{reporting machine metrics and performance}. As discussed in \cref{sec:fw_usages}, this allows for the detecting of correlations defining drill performance and improving of the smart clamp control algorithm. Monitoring of the hole and drill behaviour is also provided through images of the hole and the three-dimensional reconstruction of the drill bit for inspection.

For \textit{agent actions}, the operator may modify or begin operation of the smart clamp drilling machine. Here, we consider only those actions relevant to the information provided by the drilling machine. This includes changing of the drill bit based on the wear information, adjusting the speed of the drill or the motor pressing the drill down, or adjusting of the control algorithm for the smart clamping system.

As mentioned in \cref{sec:key_char}, there are no \textit{automatic actions} performed by the DT on the SUS. This was a conscious design choice to leave the operator in the loop. However, there are no significant technological barriers to automating some of the agent actions described here. An example of such an automation could be the automatic categorization of tool wear, and the adjustment of the drilling speed and force to account for this. This would then complete the automation loop such that the DT directly controls the SUS.

\subsection{Digital Twin Layers}
\label{sec:fw_layers}

In this section, we will break apart the black box presented in \cref{fig:dt_sus} representing the DT. As reported in many experience reports, each DT has multiple \textit{usages}, which are the reasoning activities the DT provides. In our DT description framework, we wish to break these usages apart into modular components and provide an organization such that authors can precisely report the information flow from the SUS, through the DT, and back to the SUS.

\begin{figure}[tbh]
	\centering
	\includegraphics[width=\linewidth]{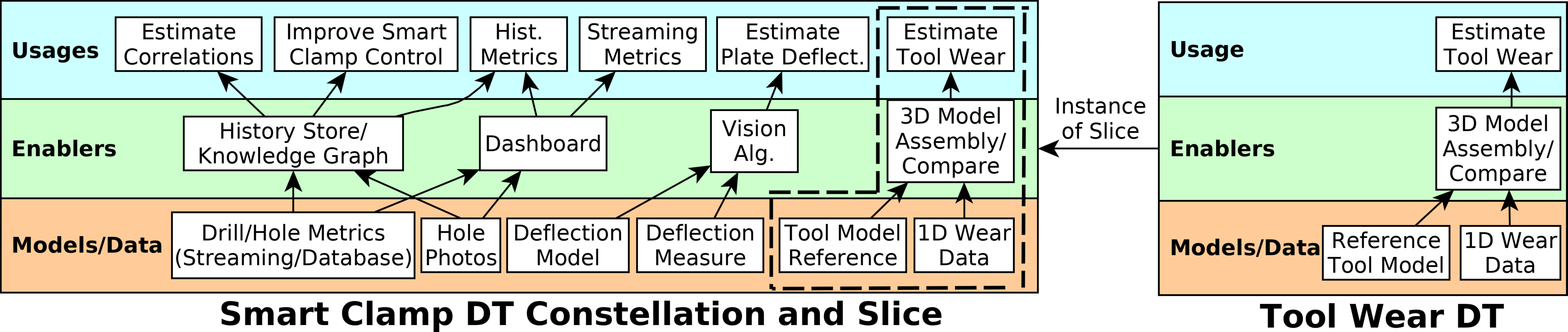}
	\caption{A layered approach to describe the smart clamp DT, and with a ``slice'' defining a DT instance.}
	\label{fig:sc_constellation}
\end{figure}

Thus, we envision three broad layers as pictured horizontally in \cref{fig:sc_constellation}: a) the \textit{usages} of a DT, b) the \textit{enabling components} which enable that usage of the DT, and c) a catch-all category for the \textit{models and data} used by the enablers. The intention with this layering is that it represents information ``flowing upwards'' through the DT gaining context and transforming from raw data into actionable insights and actions, as in the \textit{data-information-knowledge-wisdom} hierarchy~\cite{rowley2007wisdom}.

Authors of an experience report may of course wish to present a different configuration of their DT. As with other characteristics we define, this representation is coarse to offer a starting point for reporting the characteristics of a DT. These three levels were chosen to emphasize how the DT can be thought of as a collection of modular components, which we define as a \textit{DT constellation}.


\subsubsection{C7: Usages}
\label{sec:fw_usages}

One of the most crucial characteristics to understand about a DT is its \textit{usages}. That is, what activities is the DT involved in, and what direct or indirect benefits are brought to the SUS? For example, the DT may be involved in direct \textit{control and optimisation} of the processes of the SUS. Another usage could be the \textit{visualisation} of the state of the SUS, such as for design purposes, training maintenance workers, or on a dashboard. A third usage could be \textit{anomaly detection} to track the performance of the SUS and raise an error or automatically perform safety actions when the system strays outside of the safe operation range. Interested readers will find further usages in~\cite{Rasheed2020} and usages specifically for product design in~\cite{Tao2018}. In work by Govindasamy \textit{et al.}, these usages are termed the \textit{applications} of the DT~\cite{Govindasamy2021}.

As part of our research into DTs, we have also found an interesting classification of a DT's usage as either \textit{historical} or \textit{streaming/live}~\cite{Oakes2021b}. Briefly, a \textit{historical} usage is one that focuses on the past data of a SUS, while the \textit{streaming} or \textit{live} usage is one that focuses on the incoming information from a SUS. A historical usage focuses on the past information available about a SUS, such as the past behaviour or the organization's knowledge. This information is useful for usages such as detecting correlations, improving control algorithms, or referencing design iterations. In contrast, a streaming usage uses the incoming (semi-)real-time data from the SUS to provide insights or actions such as displaying a dashboard or performing command and control.

\paragraph{Smart Clamp DT:}

The smart clamp DT has a few usages as described in \cref{sec:use_case}. The historical usages include \textit{estimate correlations}, \textit{improve the smart clamp control algorithm}, and \textit{provide past drill and hole metrics}. The streaming usages include \textit{monitor of hole metrics (including a picture)}, \textit{detect plate deflection}, and \textit{estimate tool wear}.

%
%
%

\subsubsection{C8: Enablers}
\label{sec:fw_enablers}

In our conceptual layered approach visualised in \cref{fig:sc_constellation}, the \textit{enablers} rest on a layer directly below the usages. This is because the enablers are those components which take in the models and data, and in some way enable a usage. This definition for an enabler is intentionally very broad, as different domains and usages require various types of enablers.

A concrete enabler example is a state predictor and simulator which utilises machine metrics to support a predictive maintenance usage~\cite{Werner2019}. Another example would be video game engines such as Unity or Godot\footnote{\url{https://unity.com}, \url{https://godotengine.org/}} which can be used to create interactive visualisations, such as for personal health metrics~\cite{Mohammadi2020} and for training of machine operation~\cite{Karadeniz2019}.

\paragraph{Smart Clamp DT:}

A number of the smart clamp usages require the storage of historical information. This then indicates the presence of a \textit{history store} or \textit{knowledge graph} enabler storing the past metrics of the drill and the drilled holes~\cite{Lietaert2021}. Other enablers include a \textit{dashboard} for reporting metrics, a \textit{vision algorithm} for detecting plate deflection from camera input, and a \textit{tool wear estimator} to transform the one-dimensional tool dimensions into a full three-dimensional model~\cite{Bey-Temsamani2019}.

\subsubsection{C9: Models and Data}
\label{sec:fw_models_and_data}

Finally, the lowest layer of \cref{fig:sc_constellation} groups together a broad category of \textit{models and data}. This is a catch-all category for the authors of an experience report to explain which models and data are used by the DT and in particular the information flow from the SUS towards the enablers in the DT. In complex DTs, this layer may indeed be separated into different categories, such as models and data that exist in the cloud versus local storage. The relationships between the models and data layer and the enabler layer may also be bi-directional. An example is a machine learning trainer (as an enabler) which takes input data to then produce a neural net (as a model)~\cite{Min2019}.

\paragraph{Smart Clamp DT:}

The data for the smart clamp example includes everything specified in \cref{sec:fw_data_insights_actions}, such as drill and hole metrics, a measure of plate deflection, and a picture of the hole itself. The database for previous drill and hole metrics would also exist on this layer. Then, models required include a model for the vision algorithm to determine plate deflection, and tool bit reference model(s) to determine tool wear.

\subsubsection{C10: Constellations and Slices}
\label{sec:fw_constellations}

One of the most powerful aspects of DTs is that they can perform various types of reasoning about the SUS, and that DTs grow over time to support further usages as they gain enablers, models, and data~\cite{Uhlemann2017}. In our description framework, we wish to emphasize this notion of a growing DT by defining a \textit{DT constellation} as an agglomeration of related usages, enablers, models, and data for a particular SUS. This is represented visually in \cref{fig:sc_constellation} on the left-hand side, which contains multiple usages, enablers, and models/data components for the smart clamp running example.

In our view of DTs as a constellation, a myriad of connections exist between usages, enablers, models and data. That is, one usage may be supported by multiple enablers, or an enabler could use multiple models. In figures such as \cref{fig:sc_constellation}, information flowing between DT components is represented with arrows. For example, usages \textit{Historical Metrics} and \textit{Streaming Metrics} are supported by the enabler \textit{Dashboard}, which takes \textit{Streaming/Database of Drill Hole Metrics} and \textit{Hole Photos} as input.

A DT \textit{slice} is then a selection of components out of a DT constellation to support a particular usage. \Cref{fig:sc_constellation} shows one out of a possible six slices as represented by the dashed lines around the components which support usage \textit{Estimate Tool Wear}. This slice can then be implemented by any number of \textit{DT instances}, as represented in the right-hand side of \cref{fig:sc_constellation}. These slices therefore reinforce the modular and evolving nature of DTs, where the enablers and models and data are reused for multiple usages within a DT constellation and across DT projects.

Note that this representation of DT constellations and slices are conceptual objects purely for descriptive purposes. They likely are not implemented in exactly that architecture in a practitioner's DT. However, this notion of ``slicing'' out a DT instance supporting one usage assists in scoping the description of that usage's characteristics, such as insights and actions, time-scale, etc. This granularity thus allows researchers and practitioners to better understand the considerations for each DT usage.

\paragraph{Smart Clamp DT:}

On the left-hand side \cref{fig:sc_constellation}, we have reconstructed an explanatory DT constellation for the smart clamp. Across the top are the six usages described earlier, supported by the enablers and models/data described previously.

On the right-hand side of \cref{fig:sc_constellation}, we have sliced out the DT instance representing the \textit{Estimate Tool Wear} usage. This DT instance would then be receiving the \textit{one-dimensional tool wear} data, and return the insights consisting of the \textit{3D tool model comparison} to the reference model.

\subsection{Time-Scale and Fidelity Characteristics}

This section discusses important characteristics of the DT/SUS relationship: the \textit{time-scales} involved and any \textit{fidelity considerations}.

\subsubsection{C11: Time-Scales}
\label{sec:fw_time_scales}

The components of a DT are likely processing at different \textit{time-scales}. As in, data acquisition, insights, and actions can all be transmitted as \textit{slower-than-real-time} or \textit{real-time} speeds. Enablers and the usages themselves could be considered as reasoning at \textit{slower-than-real-time}, \textit{real-time}, or even \textit{faster-than-real-time} (predictive) speeds. These broad categories provide authors with the language to describe how fast their DT components are transmitting or processing.

A \textit{slower-than-real-time} scale is where communication between the SUS and the DT does not correspond to a ``live'' connection. That is, the DT periodically receives its data from the SUS, or issues insights or actions for some future time. Werner \textit{et al.}~\cite{Werner2019} provide an example for this case where data from a sensor is obtained in real-time, but the predictive maintenance insights modify worker schedules at a later time. This time-scale may also be relevant for \textit{historical} usages as described in \cref{sec:fw_usages}. These usages look at the SUS's past behaviour, as compared to the \textit{streaming} or ``real-time'' usages.

In a \textit{real-time} time-scale, data acquisition or insight/action communication is performed in a highly reactive manner, often within sub-second time differences. The intention with this time-scale category is to indicate the communication and processing which is happening in the DT/SUS relationship in a ``live'' manner, whether it is ``soft'' or ``hard'' real-time. An example of the \textit{real-time} time-scale is where the DT is directly controlling the SUS, such as real-time feedback and control in a production plant~\cite{Zhuang2018}. In such a scenario, data is gathered from the SUS, processed within the DT, and insights/actions are issued within a short amount of time, which we have referred to in earlier work as a \textit{streaming} DT~\cite{Oakes2021b}.

Finally, DTs may employ predictive simulation to
optimize the behaviour of the SUS in the (near-)future. Thus we define the time-scale of \textit{faster-than-real-time}, where enablers for a usage may predict the future trajectories of the SUS. These predictions enable either (slower-than-real-time) insights or actions like workstation layout modifications~\cite{Malik2018}, or real-time control actions like crane trajectory optimisations~\cite{Zhidchenko2018}.

As reported in our earlier work~\cite{Oakes2021}, DTs commonly involve communication and processing components at all three time-scales. The intention of introducing this characteristic is to break down how each component in the DT relates to the SUS. In particular, some domain practitioners may have the belief that ``true'' DTs are only for ``hard real-time'' control usages. That is, where the DT immediately reacts to the SUS and modifies its behaviour for optimization. Our list of characteristics is therefore designed to provide guidance for authors to discuss their visions in their experience reports. Hopefully, future research can then illustrate what the term ``digital twin'' means for each practitioner domain.

\paragraph{Smart Clamp DT:}

Most components in the smart clamp running example are soft \textit{real-time}. That is, when the drilling machine is operating, data is stored (either locally or in the cloud) and the operator sees an updated dashboard within a second or two. The exceptions to this are the usages which rely on the historical store (see \cref{fig:sc_constellation}), as finding behaviour correlations in the data and improving the smart clamp control algorithm rely on the historical data and can only be performed periodically. Therefore, these usages and components are \textit{slower-than-real-time}.

\subsubsection{C12: Fidelity Considerations}
\label{sec:fw_high_fidelity}

Worden \textit{et al.} discuss a DT as a ``mirror'' of a SUS~\cite{Worden2020} such that the DT reflects the true state and behaviour of the SUS. That is, the \textit{fidelity} between the DT and the SUS cannot be summarized as ``high'' or ``low'', but instead depends on the \textit{properties of interest} and can involve trade-offs. For example, Zhidchenko \textit{et al.} developed a DT with a simplified model to predict (in real-time) the trajectory of a mobile crane~\cite{Zhidchenko2018}. In this case the performance characteristics of the model had to be balanced against the approximation of the complex crane behaviour.

Therefore in our DT description framework we specify that authors should discuss any interesting trade-offs with respect to the fidelity of their DTs as it relates to each usage of the DT. For example, a DT usage may be visualisation for training purposes, where only coarse visual attributes such as geometry and appearance of machines are required, instead of specific details such as the surface temperature. We thus emphasize here that \textit{fidelity} refers to the DT adequately reflecting the state and behaviour of the SUS for properties relevant to each of the DTs usage. Of course, the requirements and modelling of these properties must be defined by the practitioner for their system using established modelling and engineering principles~\cite{Zeigler2000,Traore2006}.

The addition of this \textit{fidelity} characteristic to our description framework is intended to steer practitioners away from defining a complex model as \textit{high-fidelity} and therefore a ``digital twin'' of the SUS. First, what may be meant by these practitioners is that they have a \textit{digital model} as classified by Kritzinger \textit{et al.}~\cite{Kritzinger2018}. Second, it is only with respect to certain properties that a model can represent a SUS, as not every detail of a SUS can be reflected perfectly. Therefore, those properties of interest should be explicitly stated when a ``digital twin'' is created.

\paragraph{Smart Clamp DT:}

The smart clamp running example contains usages with moderate demands on fidelity. As with any manufacturing system, noise exists in the values coming from sensors. This means that the drill and hole metric values stored as history and reported on the dashboard are accurate only within some bounds. The tool wear usage is more tolerant to low fidelity insights, as tool usage degrades naturally with wear and compensation can be made for debris or containments on the monitoring system~\cite{Bey-Temsamani2019}.

\subsubsection{Digital Twin and System-under-Study Development}

These final characteristics discuss the \textit{life-cycle} of the SUS that the DT reasons about, and how the DT \textit{evolves} over the course of the project.

\subsubsection{C13: Life-cycle Stages}
\label{sec:fw_life_stages}

A SUS may have many \textit{life-cycle stages} over its existence, which may be labelled by domain-specific terms depending on whether the SUS is a manufactured product or not. For example, stages may include \textit{design}, \textit{pre-production}, and \textit{production}~\cite{Soederberg2017}, or \textit{ideation}, \textit{realisation}, and \textit{utilisation}~\cite{Siemens}. Another emerging stage is reasoning about a product's \textit{reclamation} to ease disassembly and material re-use as part of ``reverse logistics''~\cite{Pokharel2009}.


During each of these life-cycle stages, the SUS may change scope and the DTs interacting with that SUS may offer different usages. For example, during the \textit{design} stage, a DT (acting as a \textit{digital model}) may offer usages for visualising and optimising a product~\cite{Tao2019}. During the \textit{production} stage, the usages may involve each manufactured product in the manufacturing plant, such as optimising worker routines or machine settings to minimise product defects~\cite{Soederberg2017}. This could be considered an expanded SUS, or an entirely new SUS as the author prefers.

Our list of characteristics thus suggests that an author explicitly detail two aspects of their DT: a) the usages of the DT for each life-cycle stage of the SUS, and b) if the scope or structure of the SUS changes significantly throughout the stages. This assists both researchers and practitioners in understanding how the components of DTs can be utilised and re-used for each life-cycle stage. For example, enablers focusing on logistics solutions may be useful for both assembly and disassembly usages for a product.

\paragraph{Smart Clamp DT:}

For the smart clamp running example, we define two life-cycle stages: \textit{design} and \textit{operation}.

In the \textit{design} stage, the smart clamps are built and set-up. For this, the \textit{estimate correlations} usage provides the data to build and calibrate the smart clamp control algorithm. The \textit{improve smart clamp control} and \textit{historical metrics} usages are also included here, as successive versions of the smart clamp will rely on these usages to develop improved versions.

Throughout the \textit{operation} of the smart clamp drilling system, the other usages provide information to the operator. Again, the \textit{improve smart clamp control} usage could be utilised as the operator makes changes to the control algorithm. However, in the second-to-second operation of the drill, the other usages are more relevant: \textit{stream metrics}, \textit{estimate plate deflection}, and \textit{estimate tool wear}. These are the usages that the operator relies on to control the drilling process.

\subsubsection{C14: Evolution}
\label{sec:fw_evolution}

The final characteristic of our list is the \textit{evolution} of the DT throughout its development. That is, as the DT is built, connected to the SUS, and iterated upon, practitioners will encounter challenges and new requirements for the DT. As new technologies and tools are brought online to support these further usages, the DT constellation (\cref{sec:fw_constellations}) will expand to reason about more life-cycle stages or the next version of the product. For example, Soederberg \textit{et al.} report seven usages of the DT across three phases of the product life-cycle\cite{Soederberg2017}, but do not list the order in which these usages were built.

Thus the \textit{evolution} characteristic of DTs is about providing a narrative about the development of the DT as it grows and is modified throughout its development. This serves two purposes: a) connecting the usages of the DT together into a consistent story to explain the growth of the DT and the value it provides, and b) enabling further classification and research insights into this development process.

For example, it would be interesting to discover in which cases the DT project begins as a \textit{digital model}~\cite{Kritzinger2018} in a design stage, and passes through a \textit{digital shadow} stage before becoming a \textit{digital twin}. Another research question is how to organize the implementation of the DT components for a product's design-stage in parallel with the components for the product's pre-production and production stages. Finally, a third research direction is how to transfer DT components between iterations of the same product versus another product in the same family.

 \paragraph{Smart Clamp DT:}
 
 A short narrative of the development of the smart clamp and the accompanying DT has been provided in \cref{sec:use_case}. Briefly, data from the drilling machine was studied for correlations to build the smart clamp device, controlled using an algorithm built with historical data. Then, a sensor was created to detect the deflection of the plate during drilling, as well as a process for measuring the wear of the tool bit. Finally, the dashboard aspect of the project was conducted to show the operator the real-time metrics of the drilling machine.

\section{Mapping to the Asset Administration Shell}
\label{sec:aas}





Systems subject to digital twinning are used in increasingly complex architectures, often scaling across companies, designed to allow rich semantic information to flow and to be reasoned about~\cite{inigo2020towards}. Such systems encompass physical components, processes, materials, software, and other \textit{assets} that represent value to the organization~\cite{Bayha2020describing}. With this increased complexity, implementing the digital twin (DT) of the system becomes a significant challenge. To alleviate this issue, the Asset Administration Shell (AAS) has been developed by the German Electrical and Electronic Manufacturers’ Association\footnote{\url{https://www.zvei.org}} for the standardized digital representation of assets of the Reference Architecture Model Industrie 4.0 (RAMI4.0)~\cite{adolphs2015reference}. The AAS provides a machine-readable, device-independent, hierarchical standard language (metamodel) for describing the properties of assets~\cite{Bedenbender2017examples}. However, the AAS standard does not provide the means to describe the high-level capabilities of the DT. Here, we show how the AAS can be combined with our DT description framework by providing a mapping between AAS and our framework characteristics.


\subsection{Structure of the AAS}

\begin{figure}[tbp]
	\centering
    \includegraphics[width=\linewidth]{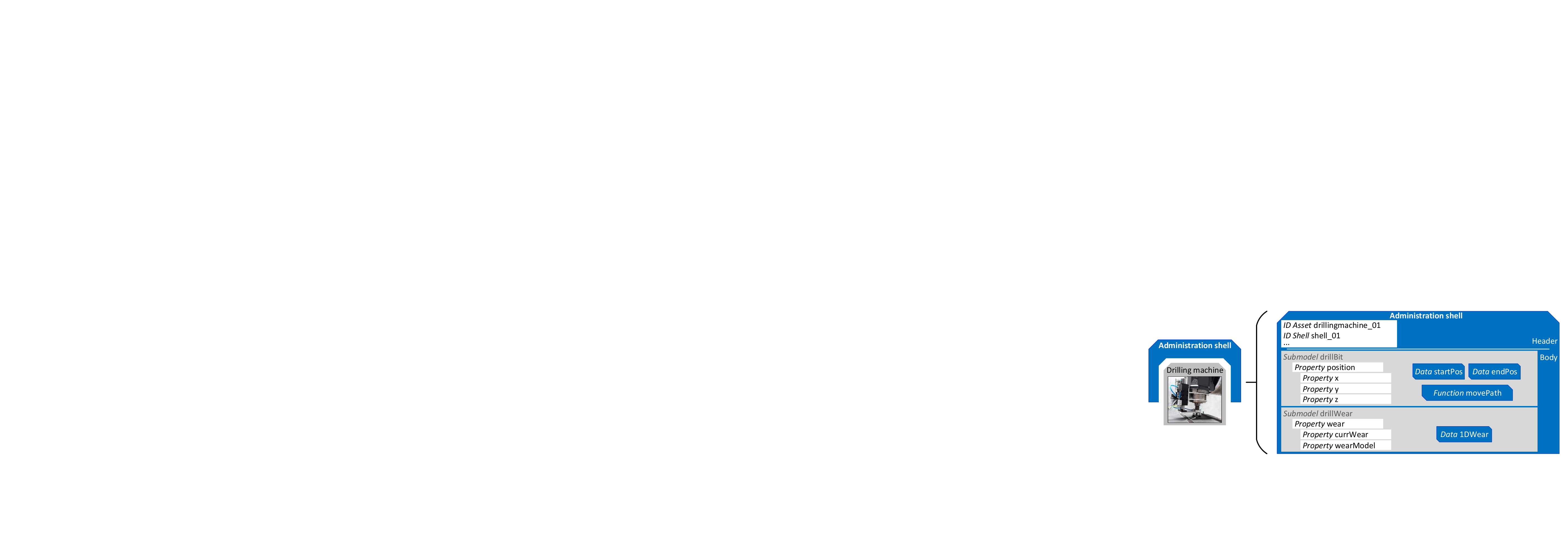}
	\caption{AAS of the drilling machine example. (Adapted from~\cite{Bayha2020describing}.)}
	\label{fig:aas_structure}
    \vspace{-10px}
\end{figure}

\Cref{fig:aas_structure} shows the (partial) structure of a possible AAS for the smart clamp drilling machine running example. The \textit{Header} contains identifying information about the assets and the shell itself. The \textit{Body} is organized into \textit{submodels} describing the asset from various views. Submodels contain a hierarchy of \textit{Properties} that are used for calculating \textit{Data}, and supporting \textit{Function}s which allow control of the asset. During operation, real-time data from the asset is directly stored within the submodels of the AAS and can be accessed through its API, allowing for structured and reusable encapsulation of assets.

For example, the submodel \textit{drillBit} captures the position of the drilling bit in a three-dimensional space; and allows reasoning about the \textit{startPosition} and \textit{endPosition} of the head during a movement action, supported by the \textit{movePath} function. The \textit{drillWear} submodel records information about the wearing of this bit, such as the incoming one-dimensional data from the sensor that is stored in the \textit{currWear} property as well as the stored three-dimensional reference model~\cite{Bey-Temsamani2019}. 

\subsection{Mapping characteristics onto the AAS}


Table~\ref{tab:aas-support} shows how the characteristics of our DT description framework are supported (or not) by the AAS. Four out of the fourteen characteristics are explicitly supported by the AAS.
That is, the AAS provides means to express and reason about these characteristics. Four characteristics are partially supported.
Four characteristics are implicitly supported,
i.e., while the structure of AAS would allow for reasoning about the characteristic, no means are provided to do so. Finally, two characteristics are not supported.

\vspace{-10px}
\begin{table}[htb!]
\caption{Support for the characteristics by the AAS}
\label{tab:aas-support}
\centering
{\footnotesize

\begin{tabular}{@{}lc@{}}
\toprule
\multicolumn{1}{c}{\textbf{Characteristic}} & \textbf{Support by the AAS} \\ \midrule
C01. System-under-study & \explicitAAS{} \\
C04. Multiplicities & \explicitAAS{} \\
C09. Models and Data & \explicitAAS{} \\
C10. Constellation & \explicitAAS{} \\
C05. Data Communicated & \partialAAS{} \\
C06. Insights and Actions & \partialAAS{} \\
C13. Life-cycle Stages & \partialAAS{} \\
C14. Evolution & \partialAAS{} \\
C02. Acting Components & \implicitAAS{} \\
C03. Sensing Components & \implicitAAS{} \\
C07. Usages & \implicitAAS{} \\
C08. Enablers & \implicitAAS{} \\
C11. Time-Scale & \nosupportAAS{} \\
C12. Fidelity Considerations & \nosupportAAS{} \\ \bottomrule
\end{tabular}
}
\caption*{\explicitAAS{}{Explicit support}\partialAAS{}{Partial support}\implicitAAS{}{Implicit}\nosupportAAS{}{No support}}
\vspace{-30px}
\end{table}


\subsubsection{Explicitly supported characteristics}

\paragraph{C1. System-under-study and C4. Multiplicities}

Practitioners can use AAS to explicitly document and scope the asset, e.g., adding text documents or representing relationships between assets and other entities. A complex asset can also be composed of other assets to explicitly represent the System-under-Study (SUS) in a hierarchical manner. For multiplicities, an AAS maps to exactly one asset, but with a complex asset the SUS can be split into multiple entities. 

\paragraph{C9. Models and Data}

Models and data themselves are represented or stored within submodels in AAS. These can be modified by incoming information from the SUS, through the AAS API, or through the functions defined on each submodel.

\paragraph{C10. Constellation}

In the AAS framework, the notion of slices where models and data flow into enablers and then usages is addressed in two ways. The first is the explicit \textit{References} between submodels and their properties. Second, \textit{Views} can provide a projection of an asset, such as showing only those AAS components relevant to a safety engineer. Above the AAS framework, slices may also be represented as the workflows in the \textit{business layer} of the RAMI 4.0 operating on the AAS components.

\subsubsection{Partially supported characteristics}

\paragraph{C5. Data Communicated}

As mentioned, AAS employs OPC-UA for communication between assets and from the physical device. However, a high-level view of what the data represents is only partially provided by typing and descriptions of the data fields. Note that AAS also specifies that multiple formats (XML, JSON, etc) can be serialised from the AAS. 

\paragraph{C6. Insights and Actions}

The AAS framework does not define the high-level insights and actions available from an asset. Instead, it mainly specifies an API to be queried by an application. For example, I{\~n}igo \textit{et al.} build a visualisation dashboard upon the data provided by their AAS of a robotic arm~\cite{inigo2020towards}. However, low-level \textit{Events} can be defined for an AAS or a submodel, representing a change of state, such as value changes.

\paragraph{C13. Life-cycle Stages}

In the RAMI 4.0, life-cycles are broken into four stages: \textit{asset type development}, \textit{asset type usage/maintenance}, \textit{asset instance production}, and \textit{asset instance usage/maintenance}. Relationships between these kinds of assets offers traceability in production and across product versions. However, the notion of usages of the DT is not explicitly connected to these life-cycle stages.

\paragraph{C14. Evolution}

Evolution of assets is performed at a low-level as asset components are versioned, and there exists a \textit{derivedFrom} relationship between AASes. This allows for tracking changes in the structure of assets. However, this does not offer the high-level narrative of the DT project as provided in Section~\ref{sec:use_case} for the smart clamp DT.

\subsubsection{Implicit characteristics}

\paragraph{C2. Acting Components and C3. Sensing Components}

In AAS, the acting and sensing components for an asset will be modelled as submodels. Communication with the physical devices is then performed using OPC-UA\footnote{\url{https://opcfoundation.org/about/opc-technologies/opc-ua/}}. However, these components are not explicitly marked and AAS does not specify whether the components were added or modified for the operation of the DT.

\paragraph{C7. Usages}

AAS provides a mechanism for defining the \textit{capabilities} of assets~\cite{Bayha2020describing}. This could be interpreted to provide a tagging system for usages denoting what reasoning an asset can provide. Alternatively, I{\~n}igo \textit{et al.} define a \textit{Condition Monitoring} submodel for their robot arm providing relevant data for visualisation~\cite{inigo2020towards}.

\paragraph{C8. Enablers}

Submodels provide \textit{operations} to provide semantics. These submodels can then represent enablers taking in models and data, and providing values to the usage submodels.

\subsubsection{Not supported characteristics}

\paragraph{C11. Time-Scale}

The notion of \textit{slower-than-real-time}, \textit{real-time}, or \textit{faster-than-real-time} is not considered in the AAS framework. Some aspects of time-scale may be handled by the OPC-UA communication layer.

\paragraph{C12. Fidelity Considerations}

The precision of data values or submodels does not seem to be considered in the AAS framework.

\subsection{Discussion}

The similarity between the structures of the AAS and our framework results in a considerable overlap in concerns. However, our framework is intended for explaining the structure and capabilities of DT in a narrative-based and rather informal way. In contrast, the structure of the AAS is explicitly defined such that technical implementations can be built for communication and control between assets. For example, our list of characteristics does not explicitly touch on technical details such as communication protocols or access control. While these are certainly crucial for DT, we leave it up to each practitioner whether to describe such mechanisms in their reports.
Due to the formal nature of the AAS, however, extending it to domains other than production and manufacturing, such as natural environment~\cite{Blair2021digital}, might be difficult. As demonstrated in our previous work~\cite{Oakes2021}, our DT description framework can be applied across multiple domains.
%
%
%
%

Characteristics such as \textit{time-scale} and \textit{fidelity considerations} are also not explicitly represented in the AAS framework, while it may be relevant to include these explicitly within a implementation framework such as AAS. Much like extensions to ontologically reason about capabilities of assets~\cite{Bayha2020describing} and combining AAS with cyber-physical systems (CPS)~\cite{Malakuti2021}, perhaps these other characteristics can be added through an extension to the AAS framework.
The explicit notion of properties in the AAS, along with the formulaic qualifiers aligns well with the notion of \textit{validity frames} for models, ensuring that models are used within their range of validity~\cite{VanAcker2021}. This improves integration efforts and can aid in selection of appropriate models for a task~\cite{VanMierlo2020}.
\section{Conclusion}
\label{sec:conclusion}

Understanding and classifying the Digital Twins (DT) described in experience reports of practitioners is essential to furthering DT research. In particular, readers should be able to easily understand the essential characteristics and capabilities of the DT at hand. The framework presented in this paper aims to improve reporting practices on DTs by defining fourteen complementary characteristics.
As an example of DT description guided by these characteristics, we describe the ``smart clamp'' drilling machine DT from \cref{sec:use_case} throughout \cref{sec:framework}. This example shows the utility of our description framework to provide clarity on both the capabilities and structure of the drilling machine DT.

Our previous report has shown that the characteristics of our description framework are relatively domain-agnostic~\cite{Oakes2021}. We hope that this assists researchers and practitioners in mapping, analysing, and comparing digital twinning practices of distant domains that would otherwise not be comparable. This DT description framework can therefore lower the barrier for knowledge- and technology- transfer across domains and aid in DT research and utilisation.


Current standardization efforts such as the Asset Administration Shell (AAS) focus on representing digital assets at technical, implementation-based level. Our DT description framework is complementary to such approaches, as it focuses on the high level capabilities of digital assets. To assess the feasibility of future integration between the two approaches, we have provided a mapping to AAS which shows a substantial overlap in structure despite the differing level of abstraction.


Future work will focus on deep integration of our framework with AAS and other frameworks, such as the Reference Architectural Model Industrie 4.0 (RAMI4.0)~\cite{adolphs2015reference}, or that of~\cite{Malakuti2021}. Such integrated approaches are sought after in advanced engineering settings, such as the Digital Thread~\cite{singh2018engineering}, e.g., for the automated extraction of a textual or visual description of the DT based on the AAS implementation, and suggesting AAS implementations based on an informal description.

\section*{ACKNOWLEDGEMENTS}

This research was supported by Flanders Make, the strategic research centre for the manufacturing industry, and was partially funded by the DTDesign ICON (Flanders Innovation \& Entrepreneurship FM/ICON :: HBC.2019.0079) project.

%
%

 \bibliographystyle{splncs04}
 \bibliography{meta/dtpaper,meta/publications}

\end{document}